\documentclass[12pt]{article}

\usepackage{amsmath}

\usepackage{amsfonts}

\usepackage{amssymb}

\usepackage[latin1]{inputenc}

\usepackage{graphics}

\usepackage{graphicx}

\usepackage{epsfig}

\usepackage{amsthm}

\usepackage{lineno}

\usepackage{mathrsfs}

\usepackage{latexsym}

\newcommand{\h}{\hspace{.5cm}}
\newenvironment{destaque}{\begin{quotation}\small\em}{\end{quotation}}

\date{}

\begin{document}

\title{Relativistic Approach to the Hydrogen Atom in a Minimal Length Scenario}
\author{{\bf R. O. Francisco, T. L. Antonacci Oakes, J. C. Fabris and}\\
 {\bf J. A. Nogueira}\footnote{e-mail: jose.nogueira@ufes.br}\\
{\small \it Departamento de F\'{\i}sica}\\
{\small \it Universidade Federal do Esp\'{\i}rito Santo}\\
{\small \it 29.075-910 - Vit\'oria-ES - Brasil}}

\maketitle

\begin{abstract}
\begin{destaque}
\h In this work we show that relativistic contributions to the ground state energy of the hydrogen atom arising from the presence of a minimal length introduced by a Lorentz-covariant algebra are more relevant than non-relativistic ones, and because of this the non-relativistic approach is not suitable. In addition, comparing our result with experimental data we can roughly estimate the upper bound for the minimal length value of the order $10^{-20}m$.
\end{destaque}
\end{abstract}

\pagenumbering{arabic}


\section{Introduction}
\label{introd}

\h The existence of a minimal length is not a new idea. Due to the divergences arising  from the advent of Quantum Field Theory, in the 1930s, Heisenberg concluded that a minimal length should exist which would be a natural cut-off for divergent integrals \cite{krag1,krag2,Hossenfelder}. In 1947, Snyder proposed a Lorentz-covariant algebra of the position and momentum operators which leads to a non-continuous space-time, and, in this way, implementing a minimal length in theory \cite{Snyder}. M. V. Battisti and S. Meljanac have analysed several physical consequences following from the noncommutative Snyder space-time \cite{battisti}.

The fact that all candidates for quantum gravity theory lead to prediction of the existence of a minimal length is not surprising, because all of them put together the fundamental constants $G$, $c$ and $\hbar$ of gravity, relativity and quantum mechanics, respectively, what allows to define a new fundamental constant,
\begin{equation}
l_{P} = \sqrt{\frac{\hbar G}{c^{3}}} \approx 1,6 \times 10^{-35} {\text m},
\end{equation}
known as Planck-length. It is clear that constructing a constant with dimension of length is not enough to assume the existence of a minimal length. A stronger argument is that the smaller the region of space-time we want to probe, the higher energy of the incident particles which we use to probe that region, consequently increasing the gravitational field created by the incident particles, and, in this way, increasingly disturbs its trajectory and increasing the uncertainty in its measurement, resulting in no measurement at all.

Since the Planck-length value is very small, the experimental researches on the existence of a minimal length at such value would not be available in now a days and would not even in close future. However, the present models of large extra dimensions have an effective scale of the Planck-length higher than four dimensions \cite{arkani, appelquist,Hossenfelder2}, what enables experimental searches for the existence of a minimal length using the current technologies \cite{das1}.

Hence, recently there is a growing interest in the search for experimental constraints to obtain an upper bound for the minimal length value. Many papers have been published concerning this issue and a wide variety of results of quite different magnitudes has been obtained.  Although most of papers are in a quantum context, S. Bensczik and collaborators have considered the effects of the classical limit (Poisson bracket) of the deformation of the canonical commutation relations on the classical orbits of particles in a central force potential \cite{benczik1}. They have found an upper bound value for the minimal length of thr order $10^{-68} m$, which is $10^{33}$ order smaller than the Planck length. In \cite{tkachuk1}, the authors have considered a treatment of many-particle which leads to an effective parameter related to the minimal length for Mercury planet inversely proportional to the number of particle of the planet and, this way, they have obtained a more satisfactory upper bound value of the minimal length of the order $10^{-17} m$.

The effects of the presence of a minimal length in the spectrum of the hydrogen atom have been calculated by many authors due to high-precision experimental measurement for the frequency of the radiation emitted during the transitions. As far as we known, Brau was who first calculated such effects and estimated a maximum value to minimal length of order $10^{-17} m$ in a 1999 paper  \cite{brau}. In 2003, R. Akhoury and Y. P. Yao, working in momentum representation, got a result different from what was obtained by Brau in position representation \cite{yao}. In 2005, S. Benczik et al also considered the spectrum of the hydrogen atom in momentum representation using perturbation theory. Their results were in agreement with Brau ones, except in the case $l = 0$, where they needed to use a numerical method and a cut-off procedure because of the divergent term $\frac{1}{r^{3}}$ \cite{benczik}. In 2006, M. M. Stetsko and V. M. Tkachuk proposed a modified perturbation theory free of divergences which allowed to calculate the corrections to all energy levels including the $l = 0$ levels \cite{stetsko}. Their results agreed with Brau ones. In 2007, K. Nouicer obtained the exact energy eigenvalues and eigenfunctions for the Coulomb potential in one dimensional space using path integral in momentum space \cite{nouicer}. In view of the discrepancy results of R. Akhoury and Y. P. Yao, in 2010, D. Bouaziz and N. Ferkous considered the same problem but proposing another method to solve the s-wave Schroendiger equation in momentum space \cite{bouaziz}. Their result for the upper bound value of the  minimal length is of the order $10^{-24} m$.

It is worth point out that in most of the papers about the hydrogen atom which are found in literature, a non-relativistic approach is used. However, we expect that relativistic effects arising from the presence of a minimal length must be more relevant than non-relativistic ones because the higher the energy, the more relevant are the effects of the existence of a minimal length.  This means that a non-relativistic approach could disregard relevant terms obtained from a relativistic approach. Therefore, it would be very interesting to study  relativistic effects on the hydrogen atom in a minimal length scenario. Thus, we propose to study the hydrogen atom  using the Dirac equation with a central potential in a minimal length scenario.

There are, however, attempts to treat the minimal length problem using Dirac equation. The authors of the reference \cite{Hossenfelder2} have obtained the Dirac equation in a minimal length scenario introduced by modifying the canonical functional relation between the momentum $\vec{p}$ and the wave vector $\vec{k}$. In order to preserve the symmetry between space and time they demand the energy $E$ and the frequency $\omega$ satisfy the same functional relation. In 2005, K. Nozari and M. Karani derived a modified Dirac equation for a free particle. They claim that due to quantum fluctuation of the background space-time, it is impossible to have free particle in a minimal length scenario \cite{nozari}. In that same year, 2005, C. Quesne and V. M. Tkachuk exactly solved the Dirac oscillator in the momentum representation \cite{tkachuk2}. One year after, in 2006, K. Nouicer published a paper where the same problem is solved \cite{nouicer1}. In 2006, C. Quesne and V. M. Tkachuk used a Lorentz-covariant algebra to deal with the (1+1)-dimensional Dirac oscillator \cite{Quesne1}. In 2011, M. I. Samar proposed a modified perturbation theory in momentum representation in order to deal with hydrogen atom in a minimal length scenario introduced by a Lorentz-covariant deformed algebra \cite{samar}. He found that the upper bound value is of the order $10^{-19} m$. In 2013, L. Menculini, O. Panella and P. Roy derived exact solutions for the (2+1)-dimensional Dirac equation in a homogeneous magnetic field \cite{panella}. In the same year, T. L. Antonacci Oakes et al calculated the energy of ground state of the hydrogen atom via Dirac equation using the Kempf's algebra \cite{oakes}. Using the non Lorentz-covariant Kempf's algebra, the Brau's result has been re-obtained.

For more about the minimal length literature, the interested reader is referred to references \cite{Mead2,chang,Sprenger,Hossenfelder}.

There are several ways of implementing a minimal length scenario. One of them is to modify the canonical commutation relations. The deformed commutation relations frequently used are ones due to Kempf \cite{Kempf2:1997}, given by\footnote{We use boldface to a vector operator for the sake of simplicity.}
\begin{equation}
	\label{rc1kempf}
	[\hat{X}_i,\hat{P}_j] = i\hbar \left[  \left(1 + \beta\hat{\textbf{P}}^2 \right) \delta_{ij} + \beta^{\prime}\hat{P}_{i}\hat{P}_{j} \right],
\end{equation}
\begin{equation}
	\label{rc2kempf}
	[\hat{X}_i,\hat{X}_j] = -i\hbar  \frac{\left[ 2\beta - \beta^{\prime} + \left(2\beta + \beta^{\prime} \right)\beta\hat{\textbf{P}}^2 \right]}{\left(1+ \beta\hat{\textbf{P}}^2 \right)}  
	\left(\hat{X}_{i}\hat{P}_{j} - \hat{X}_{j}\hat{P}_{i} \right),
\end{equation}
\begin{equation}
	\label{rc3kempf}
	[\hat{P}_i,\hat{P}_j] = 0,
\end{equation}
where $\beta$ and $\beta^{\prime}$ are parameters related to the minimal length. Nevertheless, keeping in mind that the Dirac equation is Lorentz-covariant and the algebra proposed by Kempf is not, we resort to the Lorentz-covariant algebra proposed by C. Quesne and V. M. Tkachuk \cite{Quesne1,Quesne2}, given by
\begin{equation}
	\label{quesne1}
	[\hat{X}^{\mu},\hat{P}^{\nu}] = -i\hbar \left[  \left(1 - \beta\hat{P}_{\rho}\hat{P}^{\rho} \right) g^{\mu \nu} - \beta^{\prime}\hat{P}^{\mu}\hat{P}^{\nu} \right],
\end{equation}
\begin{equation}
	\label{quesne2}
	[\hat{X}^{\mu},\hat{X}^{\nu}] = i\hbar \frac{\left[ 2\beta - \beta^{\prime} - \left(2\beta + \beta^{\prime} \right) \beta\hat{P}_{\rho}\hat{P}^{\rho} \right]} { \left(1 - \beta\hat{P}_{\rho}\hat{P}^{\rho} \right)}
 \left(\hat{P}^{\mu}\hat{X}^{\nu} - \hat{P}^{\nu}\hat{X}^{\mu} \right),
\end{equation}
\begin{equation}
	\label{quesne3}
	[\hat{P}^{\mu},\hat{P}^{\nu}] = 0,
\end{equation}
in order to introduce a minimal length in theory.

The rest of this paper is organized as follows. In section \ref{Hamls} we use a ''position'' representation, which satisfies the Lorentz-covariant commutation relations of Quesne-Tkachuk in the special case $\beta^{\prime} = 2\beta$, to modify the Dirac equation and thus to introduce the hydrogen atom in a minimal length scenario. In section \ref{Hagse} we calculate the energy of the ground state of the hydrogen atom in the minimal length scenario and roughly estimate an upper bound for the value of the minimal length. We present our conclusions in section \ref{Concl}.


\section{Hydrogen atom in a minimal length scenario}
\label{Hamls}

\h In this section we use the Lorentz-covariant deformed algebra of Quesne-Tkachuk to modify the Dirac equation with central potential in order to introduce the hydrogen atom in a minimal length scenario. We consider the case $\beta^{\prime} = 2 \beta$. Then the  equations (\ref{quesne1}), (\ref{quesne2}) and (\ref{quesne3}) to first-order of $\beta$ become
\begin{equation}
	\label{quesne1a}
	[\hat{X}^{\mu},\hat{P}^{\nu}] = -i\hbar \left[  \left(1 - \beta\hat{P}_{\rho}\hat{P}^{\rho} \right) g^{\mu \nu} - 2\beta\hat{P}^{\mu}\hat{P}^{\nu} \right],
\end{equation}
\begin{equation}
	\label{quesne2a}
	[\hat{X}^{\mu},\hat{X}^{\nu}] = 0,
\end{equation}
\begin{equation}
	\label{quesne3a}
	[\hat{P}^{\mu},\hat{P}^{\nu}] = 0.
\end{equation}
The commutation relations above lead to minimum $\Delta X^{min}_{i} = \hbar \sqrt{5 \beta}$.

It is not difficult to verify that the following non-trivial transformations of operators from $x^{\mu}$ and $p^{\mu}$ to $X^{\mu}$ and $P^{\mu}$ satisfy the Quesne-Tkachuk commutation relations (\ref{quesne1a})-(\ref{quesne3a}) to first order in $\beta$ \cite{moyedi},
\begin{equation}
\label{rx1}
	\hat{X}^{\mu} =\hat{x}^{\mu},
\end{equation}
\begin{equation}
\label{rp1}
	\hat{P}^{\mu} = \left( 1 - \beta \hat{p}^{\nu} \hat{p}_{\nu} \right) \hat{p}^{\mu},
\end{equation}
where $\hat{x}^{i}$, $\hat{p}^{i} \equiv -i\hbar \frac{\partial}{\partial x^{i}}$ ($i = 1, 2, 3$) and $\hat{p}^{0} \equiv i\hbar \frac{1}{c} \frac{\partial}{\partial t}$ are respectively position, momentum and energy operators (with the exception of c) in ordinary quantum mechanics\footnote{We use  ``ordinary quantum mechanics'' in opposition to quantum mechanics with a minimal length.}, that is, $\hat{x}^{i}$ and $\hat{p}^{i}$ satisfy
\begin{equation}
	[\hat{x}^i,\hat{x}^j] = 0,
\end{equation}
\begin{equation}
	[\hat{p}^i,\hat{p}^j] = 0,
\end{equation}
\begin{equation}
\label{rco1}
	[\hat{x}^i,\hat{p}^j] = -i\hbar g^{ij}.
\end{equation}

The Dirac equation with the electrostatic central potential of the proton in the ordinary quantum mechanics is 
\begin{equation}
\label{edirace}
\left[ - c\gamma^{0}\gamma^{\mu}\hat{p}_{\mu} + \gamma^{0}mc^{2} -\frac{\hbar c \alpha}{r} \right]| \psi(t) \rangle    = 0,
\end{equation}
where $\alpha$ is the fine structure constant and $\gamma^{\mu} \equiv \left(\hat{\beta}, \hat{\beta}\vec{\alpha} \right)$, with
\begin{equation}
\hat{\beta} = 
\begin{pmatrix}
1 & 0 \cr
0 & -1 \cr
\end{pmatrix},
\end{equation}
\begin{equation}
\vec{\alpha} = 
\begin{pmatrix}
0 & \vec{\sigma} \cr
\vec{\sigma} & 0 \cr
\end{pmatrix},
\end{equation}
and $\vec{\sigma}$ are the Pauli matrix\footnote{$\vec{\alpha}$ and $\hat{\beta}$ must be not confused with the fine structure constant $\alpha$ and the minimal length parameter $\beta$.}.

With the intention of finding out the new Dirac equation for the hydrogen atom in a minimal length scenario we replace $\hat{p}_{\mu}$ with $\hat{P}_{\mu}$ in the equation (\ref{edirace}). Because of Eq. (\ref{rx1}) the central potential is not modified in the order which we are considering, i. e., for $\cal O(\beta)$. So, the Dirac equation takes the form
$$
\left[ -i\hbar\frac{\partial}{\partial t} + c \left(\vec{\alpha} \cdot \hat{\bf p} \right) + \hat{\beta}mc^{2} -\frac{\hbar c \alpha}{r} \right] | \psi^{ML}(t) \rangle
$$
\begin{equation}
\label{rEcp}
	- \beta \left[ \frac{i \hbar^{3}}{c^{2}} \frac{\partial^{3}}{\partial t^{3}} + i\hbar \left( \vec{\alpha} \cdot \hat{\bf p} \right)^{2} \frac{\partial}{\partial t} - \frac{\hbar^{2}}{c}\left( \vec{\alpha} \cdot \hat{\bf p} \right)\frac{\partial^{2}}{\partial t^{2}} + c \left( \vec{\alpha} \cdot \hat{\bf p} \right)^{3} \right] |\psi^{ML}(t) \rangle = 0,
\end{equation}
where $\langle \psi^{ML}_{\vec{\xi}}|\psi^{ML}(t) \rangle = \psi^{ML}(\vec{\xi}, t)$ are the ''quasi-position states''\footnote{Note that $x^{i}$ is not eigenvalue of the $\hat{X}^{i}$ operator. In fact, the existence of the minimal length implies that $\hat{X}^{i}$ operator can not have any eigenstate which is a physical sate, i. e., any eigenfunction within the Hilbert space. Consequently, we are forced to introduce the so-called ''quasi-position representation'', which consists in projecting the states $|\psi^{ML}(t) \rangle $ onto the set of maximally localized states $|\psi^{ML}_{\xi} \rangle $. Thus $\langle \psi^{ML}_{\vec{\xi}}|\psi^{ML}(t) \rangle = \psi^{ML}(\vec{\xi}, t)$ are the ''quasi-position wave functions''  \cite{Kempf:1994su,Kempf1,dorsch}.}.

\section{Ground state energy}
\label{Hagse}

\h To eliminate the time in Eq. (\ref{rEcp}), we try the following ansatz
\begin{equation}
\label{ansatz}
| \psi^{ML}(t) \rangle = e^{-\frac{i}{\hbar} {\cal E} t}   | \varphi^{ML} \rangle,
\end{equation}
where $\cal E$ describes the time evolution of the stationary state $|\psi^{ML}(t)\rangle$.
Substituting (\ref{ansatz}) into (\ref{rEcp}) and neglecting terms of order ${\cal O}(\beta^{2})$, we arrive at
$$
\left[-{\cal E} + c \left(\vec{\alpha} \cdot \hat{\bf p} \right) + \hat{\beta}mc^{2} -\frac{\hbar c \alpha}{r} \right] |\varphi^{ML} \rangle +
$$
\begin{equation}
\label{rEcp1}
	\beta m^{2}c ^{2}\left[\frac{1}{m^{2}c} \left( \vec{\alpha} \cdot \hat{\bf p} \right)^{3} - \frac{{\cal E}}{m^{2}c^{2}}\left( \vec{\alpha} \cdot \hat{\bf p} \right)^{2} -
	\frac{{\cal E}^{2}}{m^{2}c^{3}}\left( \vec{\alpha} \cdot \hat{\bf p} \right) + \frac{{\cal E}^{3}}{m^{2}c^{4}} \right] | \varphi^{ML} \rangle = 0.
\end{equation}

We observe that for $\beta \rightarrow 0$, ${\cal E}$ is the ordinary energy  $E$ of the hydrogen atom. Therefore, if we assume the mass scale of the minimal length $M_{ML}$ to be big so that the electron mass is much smaller than it ($\beta = \frac{c^{2}}{M^{2}_{ML}c^{4}}$, so $\beta m^{2}c^{2} = \frac{m^{2}}{M^{2}_{ML}} \ll 1$), then we can consider the second term as a perturbation. Consequently, Eq. (\ref{rEcp1}) suggests we may assume that
\begin{equation}
\label{aprox1}
{\cal E} = E^{ML} = E + \beta m^{2}c^{2}E^{1} + {\cal O}(\beta^{2})
\end{equation}
and
\begin{equation}
\label{aprox2}
 |\varphi^{ML} \rangle = |\varphi \rangle + \beta m^{2}c^{2} |\varphi^{1} \rangle + {\cal O}(\beta^{2}),
\end{equation}
where $E$ is the energy of the eigenstate $|\varphi\rangle$ of the hydrogen atom obtained from the ordinary Dirac equation.

Substituting (\ref{aprox1}) and (\ref{aprox2}) into (\ref{rEcp1}) and neglecting terms of order ${\cal O}(\beta^{2})$, we obtain
\begin{equation}
\label{energiaml}
E^{ML} = E + \beta \langle \varphi | \left[c\left( \vec{\alpha} \cdot \hat{\bf p} \right)^{3} - E \left( \vec{\alpha} \cdot \hat{\bf p} \right)^{2} -
	\frac{E^{2}}{c}\left( \vec{\alpha} \cdot \hat{\bf p} \right) + \frac{E^{3}}{c^{2}} \right]| \varphi \rangle.
\end{equation}

Although, in general the calculation of the expression above is very laborious, it can be performed without difficulty in the case of the ground state. Then, from Eq. (\ref{energiaml}) we have
\begin{equation}
\label{eml}
E_{0}^{ML} = E_{0} + \beta \left[c\langle \varphi_{0} | \left( \vec{\alpha} \cdot \hat{\bf p} \right)^{3} | \varphi_{0} \rangle
- E_{0} \langle \varphi_{0} | \left( \vec{\alpha} \cdot \hat{\bf p} \right)^{2} | \varphi_{0} \rangle
- \frac{E_{0}^{2}}{c} \langle \varphi_{0} | \left( \vec{\alpha} \cdot \hat{\bf p} \right) | \varphi_{0} \rangle 
+ \frac{E_{0}^{3}}{c^{2}} \right],
\end{equation}
where $E_{0} = mc^{2}\sqrt{1-\alpha^{2}}$ is the energy of the ground state of the hydrogen atom, $ | \varphi_{0} \rangle $, obtained from the ordinary Dirac equation.  

Thereby, we need to calculate the expressions
\begin{equation}
\langle \varphi_{0} | \left( \vec{\alpha} \cdot \hat{\bf p} \right) | \varphi_{0} \rangle =
\int  \left(\phi^{\dag}_{0}, \chi^{\dag}_{0} \right)
\begin{pmatrix}
0 & \vec{\sigma} \cdot \hat{\bf p} \cr
\vec{\sigma} \cdot \hat{\bf p} & 0 \cr
\end{pmatrix}
\begin{pmatrix}
\phi_{0} \cr
\chi_{0} \cr
\end{pmatrix}
d^{3}\vec{x},
\end{equation}

\begin{equation}
\langle \varphi_{0} | \left( \vec{\alpha} \cdot \hat{\bf p} \right)^{2} | \varphi_{0} \rangle =
\int  \left(\phi^{\dag}_{0}, \chi^{\dag}_{0} \right)
\begin{pmatrix}
\left( \vec{\sigma} \cdot \hat{\bf p} \right)^{2} & 0 \cr
0 & \left( \vec{\sigma} \cdot \hat{\bf p} \right)^{2} \cr
\end{pmatrix}
\begin{pmatrix}
\phi_{0} \cr
\chi_{0} \cr
\end{pmatrix}
d^{3}\vec{x},
\end{equation}

\begin{equation}
\langle \varphi_{0} | \left( \vec{\alpha} \cdot \hat{\bf p} \right)^{3} | \varphi_{0} \rangle =
\int  \left(\phi^{\dag}_{0}, \chi^{\dag}_{0} \right)
\begin{pmatrix}
0 & \left( \vec{\sigma} \cdot \hat{\bf p} \right)^{3} \cr
\left( \vec{\sigma} \cdot \hat{\bf p} \right)^{3} & 0 \cr
\end{pmatrix}
\begin{pmatrix}
\phi_{0} \cr
\chi_{0} \cr
\end{pmatrix}
d^{3}\vec{x}.
\end{equation}

The two-component eignspinors of the ground state, $\phi_{0}$ and $\chi_{0}$, are given by
\begin{equation}
\langle \vec{x} | \varphi_{0} \rangle =
\begin{pmatrix}
\phi_{0} \cr
\chi_{0} \cr
\end{pmatrix}
=
\begin{pmatrix}
F(r)Y^{1/2, m}_{0} \left(\theta, \phi \right) \cr
-if(r)Y^{1/2, m}_{1} \left(\theta, \phi \right) \cr
\end{pmatrix},
\end{equation}
where
\begin{equation}
F(r) = a_{0}b r^{\gamma} e^{-ar},
\end{equation}
\begin{equation}
f(r) = b_{0}b r^{\gamma} e^{-ar},
\end{equation}
with
\begin{equation}
\gamma = \epsilon - 1,
\end{equation}
\begin{equation}
a =  \left( \frac{mc}{\hbar} \right)\sqrt{1 - \epsilon^{2}},
\end{equation}
\begin{equation}
b = \left( \frac{mc}{\hbar} \right)^{\gamma},
\end{equation}
\begin{equation}
a_{0} = \left( \frac{2a}{b} \right)^{\gamma + 1} \sqrt{\frac{\left(1 + \epsilon \right)}{\Gamma \left( 2\gamma +3 \right) }},
\end{equation}
\begin{equation}
b_{0} = \sqrt{\frac{1 - \epsilon}{1 + \epsilon}}a_{0},
\end{equation}
\begin{equation}
 \epsilon = \frac{E_{0}}{mc^{2}},
\end{equation}
and $ Y^{j, m}_{j \pm 1/2} \left(\theta, \phi \right) $ are the common eigenspinor-function of $\hat{j}_{z}$ and $\hat{J}^{2}$ \cite{Merzbacher}.

Now, we employ the following identity
\begin{equation}
\label{ident1}
\vec{\sigma} \cdot \hat{\bf p} = \vec{\sigma} \cdot \vec{e}_{r} \left( -i\hbar \frac{\partial}{\partial r} + i\frac{\vec{\sigma} \cdot \hat{\bf L}}{r} \right),
\end{equation}
with
\begin{equation}
\label{ident2}
\vec{\sigma} \cdot \vec{e}_{r}Y^{j, m}_{j \pm 1/2}  = - Y^{j, m}_{j \pm 1/2},
\end{equation}
in order to get
\begin{equation}
\label{1fi1}
\left( \vec{\sigma} \cdot \hat{\bf p} \right) \phi_{0} = i\hbar \frac{dF}{dr}Y^{1/2, m}_{1},
\end{equation}
\begin{equation}
\label{1fi2}
\left( \vec{\sigma} \cdot \hat{\bf p} \right) \chi_{0} = \left( \hbar \frac{df}{dr} + 2\hbar \frac{f}{r} \right)Y^{1/2, m}_{0},
\end{equation}
and
\begin{equation}
\label{2fi1}
\left( \vec{\sigma} \cdot \hat{\bf p} \right)^{2} \phi_{0} = -\hbar^{2} \left( \frac{d^{2}F}{dr^{2}} + 2\frac{1}{r}\frac{dF}{dr} \right)Y^{1/2, m}_{0},
\end{equation}
\begin{equation}
\label{2fi2}
\left( \vec{\sigma} \cdot \hat{\bf p} \right)^{2} \chi_{0} = i\hbar^{2} \left( \frac{d^{2}f}{dr^{2}} + 2\frac{1}{r}\frac{df}{dr} -2\frac{f}{r^{2}} \right)Y^{1/2, m}_{1},
\end{equation}
and
\begin{equation}
\label{3fi1}
\left( \vec{\sigma} \cdot \hat{\bf p} \right)^{3} \phi_{0} = - i\hbar^{3} \left( \frac{d^{3}F}{dr^{3}} + 2 \frac{1}{r}\frac{d^{2}F}{dr^{2}} -2\frac{1}{r^{2}}\frac{dF}{dr} \right) Y^{1/2, m}_{1},
\end{equation}
\begin{equation}
\label{3fi2}
\left( \vec{\sigma} \cdot \hat{\bf p} \right)^{3} \chi_{0} = -\hbar^{3} \left( \frac{d^{3}f}{dr^{3}} + 4 \frac{1}{r}\frac{d^{2}f}{dr^{2}} \right)Y^{1/2, m}_{0}.
\end{equation}

After some algebra we have
\begin{equation}
\langle \varphi_{0} | \left( \vec{\alpha} \cdot \hat{\bf p} \right) | \varphi_{0} \rangle = \frac{mc}{\epsilon}\left( 1- \epsilon^{2} \right),
\end{equation}
\begin{equation}
\langle \varphi_{0} | \left( \vec{\alpha} \cdot \hat{\bf p} \right)^{2} | \varphi_{0} \rangle =
 m^{2}c^{2} \frac{\left(2 - \epsilon \right) \left( 1- \epsilon^{2} \right)}{\epsilon  \left(2\epsilon -1 \right)},
\end{equation}
and
\begin{equation}
\langle \varphi_{0} | \left( \vec{\alpha} \cdot \hat{\bf p} \right)^{3} | \varphi_{0} \rangle =
 m^{3}c^{3}\frac{\left( 1- \epsilon^{2} \right)^{2}}{\epsilon \left(2 \epsilon - 1 \right)}.
\end{equation}

At last, after more calculations, we find
\begin{equation}
\label{er}
E_{0}^{ML} = mc^{2}\epsilon + \beta m^{3}c^{4} \left( \frac{1 - 2\epsilon - 2\epsilon^{4} + 4\epsilon^{5} }{2\epsilon^{2} - \epsilon} \right).
\end{equation}

It is interesting to expand $E_{0}^{ML}$ in power of the fine structure constant. After performing some simple calculations, we get
\begin{equation}
\label{edml}
E_{0}^{ML} \approx  mc^{2} \left( 1 - \frac{\alpha^{2}}{2} - \frac{\alpha^{4}}{8} \right) +
 \beta m^{3}c^{4} \left(1 - \frac{7\alpha^{2}}{2} + \frac{3\alpha^{4}}{8} \right).
\end{equation}

As it is clearly seen, the summation of the terms independent of the fine structure constant in Eq. (\ref{edml}) is the electron rest energy (what is in agreement with \cite{moyedi}). Hence, subtracting the rest energy of the electron,
\begin{equation}
\Delta E_{0}^{ML} = E_{0}^{ML} - mc^{2} - \beta m^{3}c^{4},
\end{equation}
we get
\begin{equation}
\label{edm2}
\Delta E_{0}^{ML} \approx - mc^{2} \left( \frac{\alpha^{2}}{2} + \frac{\alpha^{4}}{8} \right) -  
\beta m^{3}c^{4} \left( \frac{7\alpha^{2}}{2} - \frac{3\alpha^{4}}{8} \right).
\end{equation}

This result shows that the correction to the energy of the ground state of the hydrogen atom is always negative and of ${\cal O}(\alpha^{2})$, which is in agreement with the reference \cite{samar}.

Naively we could expect that in the limit small $\alpha$ we would recover the Brau's result \cite{brau}. However, as we suspected, a careful examination clearly reveals that the relativistic effects stem from the presence of a minimal length introduced by Lorentz-covariant algebra start at order ${\cal O}(\alpha^{2})$, instead of ${\cal O}(\alpha^{4})$ as in the works with non-relativistic approach. In conclusion, when one considers relativistic effects as less important, more relevant terms may be lost.

We can make a estimation of the minimal length value comparing our theoretical result with the experimental data of the measurement of the 1S-2S energy splitting in the hydrogen atom. As far as we know, the best accuracy has been obtained by C. G. Parthey et al. \cite{parthey}. They have gotten an accuracy of 4 parts in $10^{15}$ (2,466,061,413,187,035(10)Hz). Indeed, we can make a rough estimative of the maximum value of the minimal length, if we realize the contribution of the lowest order term to the correction of the energy of the 2S state arising from the presence of a minimal length must be of ${\cal O}(\alpha^{2})$, because the 1S and 2S ordinary states have the same symmetry. Therefore, if this error is entirely attributed to the minimal length corrections and we assume that the effects of the minimal length can not yet be seen experimentally, from (\ref{edm2}) we find
\begin{equation}
\Delta X_{i}^{min} \leq 10^{-20} m.
\end{equation}


\section{Sumary and Conclusion}
\label{Concl}

\h The aim of this work was to show that the relativistic contributions to the ground state of the hydrogen atom arising from the presence of a minimal length are more relevant than non-relativistic ones. The hydrogen atom have been introduced in a minimal length scenario by modifying the Dirac equation with central potential through the use of the Lorentz-covariant deformed algebra of Quesne-Tkachuk and in the special case $\beta^{\prime} = 2\beta$, see Eqs. (\ref{quesne1a}), (\ref{quesne2a}) and (\ref{quesne3a}). In order to avoid the problem of substituting $\hat{X}_{i}$ for derivatives of $\hat{p}_{i}$ in the Coulomb potential ($\frac{1}{r}$) we have used the ``position'' representation given by equations (\ref{rx1}) and (\ref{rp1}). Assuming that the electron mass is much smaller than the mass scale of the minimal length, we can calculate the energy shift of the ground state of the hydrogen atom in a perturbative way. 

By expanding the ground state energy of the hydrogen atom, see Eq. (\ref{er}), in power of the fine structure constant, we have found that the energy shift is of ${\cal O}(\alpha^{2})$, consequently two orders lower than one found by Brau. Therefore, in agreement with our statement that relativistic effects of the presence of a minimal length are more relevant than non-relativistic ones.

It is interesting to point out that if instead of a Lorentz-covariant algebra we consider the Kempf algebra \cite{oakes}, see Eqs (\ref{rc1kempf}), (\ref{rc2kempf}) and (\ref{rc3kempf}), the Eq.(\ref{eml}) becomes
\begin{equation}
\label{em2}
E_{0}^{ML} = E_{0} + \beta c\langle \varphi_{0} | \left( \vec{\alpha} \cdot \hat{\bf p} \right)^{3} | \varphi_{0} \rangle ,
\end{equation}
for which the ground state energy is given by 
\begin{equation}
\label{edmc1}
E_{0}^{ML} \approx  mc^{2} \left( 1 - \frac{\alpha^{2}}{2} - \frac{\alpha^{4}}{8} \right) + \beta m^{3}c^{4} \alpha^{4}.
\end{equation}
The above result shows that the correction of lowest order due to the presence of a minimal length in this case is of same order as non-relativistic one. This means that the use of the Kempf algebra, which is not Lorentz-covariant, provides terms of the same relevance in both relativistic and non-relativistic approaches.  However the use of the Kempf algebra does not leave the Dirac equation manifestly symmetric in space and time: the treatment of space and time at same level can be recovered only assuming ad-hoc a modification in the canonical functional relation between the energy operator and the generator of time translation \cite{Hossenfelder2,oakes}.

It is important emphasize that it is necessary to reconsider the Lorentz covariance in the presence of a minimal length. Recently, A. F. Ali, S. Das and E. C. Vagenas proposed a generalization of the uncertainty principle (GUP) \cite{ali1,ali2} which is consistent with special relativity theories  (Doubly Special Relativity) and include a minimal length as fundamental limit for contractions of space. In the modified Dirac equation resulting from the GUP proposed by them we can find the $\left( \vec{\alpha} \cdot \hat{\bf p} \right)^{2}$ term, which leads to terms of ${\cal O}(\alpha^2)$ in the ground state energy.

At last, comparing our result with experimental data we can roughly estimate the upper bound for the minimal length value of the order of $10^{-20}m$. It is important to point out that the length scale of the Large Hadron Collider (LHC)is of order of $10^{-19} m$, therefore very close to ours and lower than the Brau's one.


\section*{Acknowledgements}

\h We would like to thank FAPES, CAPES and CNPq (Brazil) for financial support.
\\
\\
\\



\end{document}